\documentclass[pra,aps,twocolumn,superscriptaddress,floatfix,showpacs,longbibliography]{revtex4-2}
\usepackage{graphicx}
\usepackage{dcolumn}
\usepackage{bm}
\usepackage{upgreek}
\usepackage{epsfig}
\usepackage{epstopdf}
\usepackage{natbib}
\usepackage[normalem]{ulem}
\usepackage{amsmath,amssymb}
\usepackage{dsfont}
\usepackage[english]{babel}
\usepackage{braket}
\usepackage{color}

\begin{document}

\title{Induced Interactions and Bipolarons in Spin-Orbit Coupled Bose-Einstein Condensates}

\author{Zhe Yang}
\affiliation{Guangdong Basic Research Center of Excellence for Structure and Fundamental Interactions of Matter, Guangdong Provincial Key Laboratory of Quantum Engineering and Quantum Materials, School of Physics, South China Normal University, Guangzhou 510006, China}
\affiliation{Guangdong-Hong Kong Joint Laboratory of Quantum Matter, Frontier
Research Institute for Physics, South China Normal University, Guangzhou
510006, China}
\author{Shanshan Ding}
\affiliation{Institute of Atomic and Molecular Physics, Sichuan University, Chengdu 610065, China}
\author{Qizhong Zhu}
\email{qzzhu@m.scnu.edu.cn}
\affiliation{Guangdong Basic Research Center of Excellence for Structure and Fundamental Interactions of Matter, Guangdong Provincial Key Laboratory of Quantum Engineering and Quantum Materials, School of Physics, South China Normal University, Guangzhou 510006, China}
\affiliation{Guangdong-Hong Kong Joint Laboratory of Quantum Matter, Frontier
Research Institute for Physics, South China Normal University, Guangzhou
510006, China}

\date{\today}

\begin{abstract}
Impurities immersed in a Bose-Einstein condensate (BEC) can interact indirectly through the exchange of Bogoliubov excitations. These impurities, which form dressed quasiparticles known as Bose polarons due to their interaction with the BEC, can pair up to form a bound state called bipolarons, via an induced interaction. Previous studies on induced interactions have primarily focused on cases with an isotropic excitation spectrum. In this work, we investigate the properties of induced interactions and bipolarons mediated by anisotropic Bogoliubov excitations using field theory. 
Taking a BEC with spin-orbit coupling as an example, we show that the induced interaction becomes anisotropic. Notably, a double-minima feature appears in the induced interaction in momentum space due to the exchange of roton excitations. 
Additionally, we calculate the binding energy and wave functions of these bipolarons induced by anisotropic interactions. Unlike previously studied bipolarons formed through the exchange of isotropic phonon excitations, we identify a new type of bipolarons whose wave functions feature a double-peak structure under strong impurity-boson interactions. Our work extends the theory of induced interactions from isotropic to anisotropic systems, and
reveals the novel features in both the induced interactions and bipolarons arising from BEC with an unconventional excitation spectrum.
\end{abstract}

{\maketitle}

\section{introduction}
Quasiparticle is an essential concept for describing the effective behavior of complex many-body systems, as it simplifies analysis by providing a clear and intuitive physical framework \cite{baymlandau2004}. It is common in condensed matter systems and include examples such as quasiparticles in Fermi liquids, phonons in solids, excitons in semiconductors, exciton-polaritons in semiconductor cavities and polarons. The concept of polaron was first proposed by Landau as a quasiparticle describing the electron interacting with and dressed by lattice phonons. In ultracold atomic gases, impurities immersed in bosonic or fermionic gases enable the realization of Bose or Fermi polarons. Ultracold atom platforms offer significant advantages, providing a systematic and highly controllable approach to studying polarons. In experiments,
both Bose and Fermi polarons are successfully observed
\cite{schirotzek2009observation,kohstall2012metastability,scazza2017repulsive,cetina2016ultrafast,jorgensen2016observation,hu2016bose,desalvo2019observation,yan2020bose,nascimbene2009collective}, leading to extensive theoretical investigations into Fermi polarons \cite{hu2018attractive,scazza2022repulsive,tajima2021polaron,massignan2014polarons,yi2015,chen2016,peng2021,deng2018,hu2016} and Bose polarons \cite{guenther2018bose,li2014variational,levinsen2015impurity,christensen2015quasiparticle,shchadilova2016quantum,tempere2009feynman,grusdt2017strong,ardila2015impurity,ardila2016bose,sun2017}.

Induced interactions between quasiparticles are widely observed across various fields of physics, as comprehensively reviewed in a recent work by Paredes et al. \cite{paredes_interactions_2024}. For example, in conventional superconductors, electron pairing occurs due to the exchange of lattice phonons. For Bose polarons in cold atoms, the induced interaction between impurities are brought by the exchange of low-energy excitations of the Bose-Einstein condensate (BEC) medium. The induced interaction is overall attractive and may lead to the formation of polaron bound states known as bipolarons. 
So far, the induced interaction from isotropic Bogoliubov excitations has been extensively studied in both theory and experiment \cite{yu2012induced,fujii2022universal,he2017Casimir,drescher2023medium,klein2005interaction,naidon2018two,camacho2018bipolarons,camacho2018landau,dehkharghani2018coalescence,Kinnunen2018,Schecter2014}. It is natural to ask how the induced interaction arising from anisotropic Bogoliubov excitations differs from the conventional case. Furthermore, could new types of bipolarons form with characteristics different from conventional ones?

In this paper, we study the properties of induced interaction and bipolarons mediated by anisotropic Bogoliubov excitations in BEC, taking BEC with spin-orbit coupling (SOC) as an example \cite{lin2011spin,wang2012spin,wang2010spin,li2012quantum,li2017stripe,ji2015softening,ho2011bose,martone2023bose,chen2018quantum,zhu2012exotic,Zhai_2015}. It is well known that BEC with SOC has anisotropic excitations dependent on the form of SOC. In particular, we consider a one-dimensional (1D) SOC with equal Rashba and Dresselhaus type which has been realized in experiments and studied extensively \cite{lin2011spin,Zhai_2015}. The excitation spectrum features a phonon dispersion at small momentum and roton dispersion at finite momentum. Previously, it has been found that Bose polaron in this system has novel features such as polaron dispersion miminum at nonzero momentum \cite{Huhui2019}. However, the induced interaction between these polarons has not yet been investigated. Here, we find that the induced interaction becomes anisotropic as a result of the anisotropic Bogoliubov excitations. Additionally, the roton dispersion leads to a double-minima structure for the induced interaction in momentum space. The induced interaction is calculated with both field theory and second-order perturbation theory, where the former method is applicable even for strong impurity-boson interaction. We further calculate the binding energy and wave functions of bipolarons, and find that for strong impurity-boson interaction the bipolaron wave functions feature a novel double-peak structure in momentum space.

The structure of the rest of this paper is organized as follows. In Sec. \ref{soc_bec}, we provide a brief review of Bogoliubov excitations in spin-orbit coupled BEC using the method of band projection, i.e., by projecting the Hamiltonian onto the lower band before performing mean-field approximation. In Sec. \ref{interaction}, we explore the properties of the Bose polaron in a spin-orbit coupled BEC medium and calculate the induced interactions mediated by anisotropic Bogoliubov excitations using field theory, which remains valid even for strong impurity-boson interactions. In particular, we find that the roton-induced interaction has a double-minima feature in momentum space. In Sec. \ref{bipolaron}, we examine the properties of bipolarons arising from the anisotropic interaction and reveal the feature  of bipolarons resulting from anistropic interactions. Finally, Sec. \ref{conclusion} concludes this paper with a comprehensive summary of our key findings and results.

\section{excitations in spin-orbit coupled BEC}
\label{soc_bec}
BECs with anisotropic excitation spectra are commonly found in systems confined within anisotropic optical lattices or those with broken time-reversal symmetry. Here, we take spin-orbit coupled BECs as an example. These systems exhibit an anisotropic excitation spectrum, which differs between the direction influenced by SOC and the perpendicular direction. Furthermore, for the plane wave phase where the BEC condenses at nonzero momentum, the spectrum becomes asymmetric with respect to inversion along the SOC direction. Spin-orbit coupled bosonic system serves as an example to examine how anisotropic and asymmetric dispersions impact induced interactions.

To be specific, consider a BEC with Raman-induced SOC in the $x$ direction, whose single-particle Hamiltonian of the system reads (set $\hbar$=1) \cite{Zhaihui2013,zhai2012spin},
\begin{equation}
H_{0}=\sum_{\boldsymbol{k}}\left(a_{\boldsymbol{k}\uparrow}^{\dagger}, a_{\boldsymbol{k}\downarrow}^{\dagger}\right)\left[\dfrac{(\boldsymbol{k}-k_{0}\boldsymbol{e}_{x}\sigma_{z})^{2}}{2m_{B}}+\dfrac{\Omega\sigma_{x}}{2}\right]
\left(\begin{array}{c}
a_{\boldsymbol{k}\uparrow}\\
a_{\boldsymbol{k}\downarrow}
\end{array}\right),
\end{equation}
where $m_B$ is the boson atomic mass, $k_{0}$ is the recoil momentum, $\Omega$  is the Raman coupling strength, and $a_{k\sigma}$ is the annihilation operator for a pseudospin state $\sigma=\{\uparrow,\downarrow\}$ at momentum $\boldsymbol{k}$. $\sigma_{x}$ and $\sigma_{z}$ are the $x$ and $z$ components of Pauli matrices, respectively. Hereafter, we set $k_{0}$ as the momentum unit and the associated recoil energy $E_{r}={k_{0}^{2}}/{2m_{B}}$ as the energy unit. The interaction term is 
\begin{equation}
H_{\textrm{int}}=\dfrac{g_{\sigma\sigma^{\prime}}}{2V}\sum_{\boldsymbol{k_{1}k_{2}q}}a_{\boldsymbol{k_{1}+q}\sigma}^{\dagger}a_{\boldsymbol{k_{2}-q}\sigma^{\prime}}^{\dagger}a_{\boldsymbol{k_{2}}\sigma^{\prime}}a_{\boldsymbol{k_{1}}\sigma},
\end{equation}
where for simplicity, the SU(2)-invariant interactions between bosons are considered, i.e., $g_{\sigma\sigma^{\prime}}=g={4\pi a_{B}}/{m_{B}}$, with $a_B$ being the boson-boson scattering length.

\begin{figure}[t]
	\centering
	\includegraphics[width=0.98\linewidth]{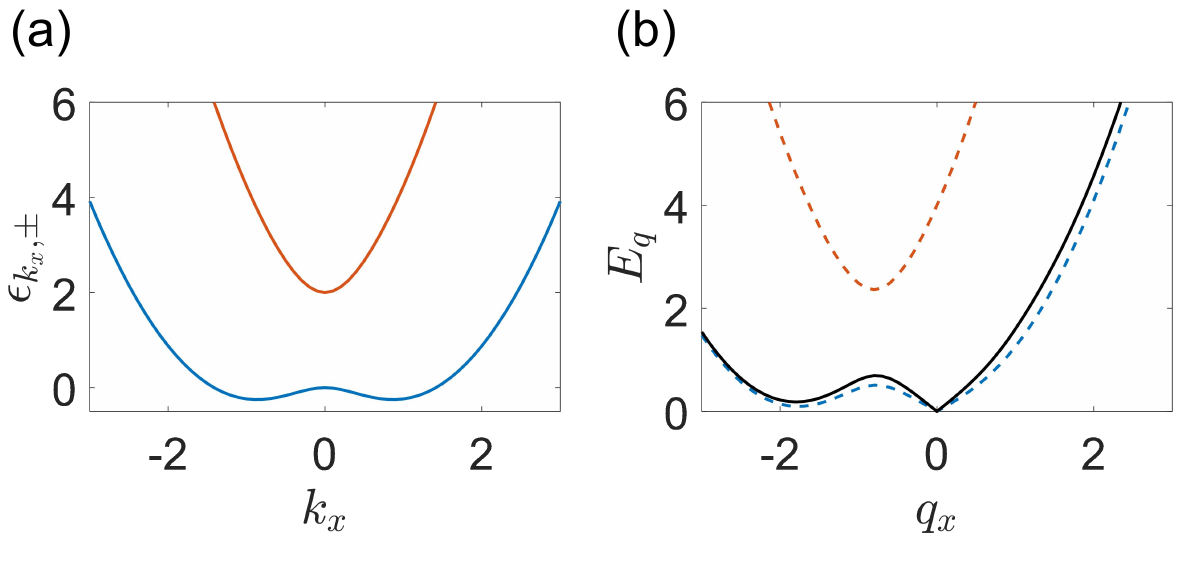}
  \caption{(a) The single particle dispersion $\epsilon_{\boldsymbol{k}, \pm}$ along the $k_{x}$ direction (setting $k_y=0$). (b) The Bogoliubov excitation $E_{\boldsymbol{q}}$ along the $q_x$ direction (setting $q_y=0$). The dashed lines are the full spectrum and the solid line is the one after band projection. The parameters chosen are $\Omega/E_{r}=2$, $n_{0}=1/k^{3}_{r}$, $gn_{0}/E_{r}=0.5$,
and we will fix these three parameters in the following calculation.  }
    \label{fig1}
\end{figure}

The energy spectrum of the single-particle Hamiltonian $H_{0}$ has two branches, corresponding to two helicities,
\begin{equation}
 \epsilon_{\boldsymbol{k}, \pm}=\frac{k^{2}}{2 m_B}+\frac{k_{0}^{2}}{2 m_B} \pm \sqrt{\left(\frac{k_{x} k_{0}}{m_B}\right)^{2}+\left(\frac{\Omega}{2}\right)^{2}},   
\end{equation}
where $k=\sqrt{k_x^2+k_y^2+k_z^2}$. When $\Omega<4E_{r}$, the lower branch $ \epsilon_{\boldsymbol{k},-}$ has two minima at $\pm\boldsymbol{k}_{m}=\pm k_{0}\cos2\theta\boldsymbol{e}_{x}$ with $2\theta=\arcsin(\Omega/4E_{r})$. In this case the ground state of spin-orbit coupled BEC is a plane wave phase with bosons condensed at either of the two minima as show in Fig. \ref{fig1}, and the low energy excitation spectra features an asymmetric dispersion in the $x$ direction. The focus of this study is on the properties of bipolarons induced by this type of anisotropic and asymmetric Bogoliubov excitation.

The Bogoliubov excitation of this system has being extensively studied \cite{Zhaihui2013,martone2023bose}, and the excitation has two branches, due to the pseudospin-1/2 character of boson, as shown in Fig. \ref{fig1}. The two branches are separated by an energy gap, and thus at very low temperature, we expect that only the lower branch will have major impact on the induced interaction. So for simplicity, we first project the single-particle Hamiltonian to the lower branch, and then obtain the lower branch of Bogoliubov excitation through mean-field theory. This projection method dramatically simplifies the following calculations while preserving the essential physics.


To be specific, we first expand the operator $a_{\boldsymbol{k}\sigma}$ in the eigenstates of $H_0$, i.e., $a_{\boldsymbol{k}\sigma}=\langle\boldsymbol{k}\sigma|\boldsymbol{k}+\rangle a_{\boldsymbol{k}+}+\langle\boldsymbol{k}\sigma|\boldsymbol{k}-\rangle a_{\boldsymbol{k}-}$, where $\langle\boldsymbol{k}\sigma|\boldsymbol{k}\pm\rangle$ is the transformation matrix which diagonalizes $H_0$. By discarding the occupation of the upper energy branch, $a_{\boldsymbol{k}\sigma}\approx\langle\boldsymbol{k}\sigma|\boldsymbol{k}-\rangle a_{\boldsymbol{k}-}$, we obtain the projected Hamiltonian with interaction \cite{subacsi2022quantum,frolian2022realizing,cui2013}
\begin{equation}
\begin{split}
H_{\text{BEC}}  \simeq & \sum_{\boldsymbol{k}}\epsilon_{\boldsymbol{k},-} a_{\boldsymbol{k},-}^{\dagger}a_{\boldsymbol{k},-}\\
      & +\dfrac{1}{2V}\sum^{\prime}
      f_{\boldsymbol{k_{1},k_{2}}}^{\boldsymbol{k_{3},k_{4}}}
      a_{\boldsymbol{k_{1}},-}^{\dagger}a_{\boldsymbol{k_{2}},-}^{\dagger}
      a_{\boldsymbol{k_{3}},-}a_{\boldsymbol{k_{4}},-}.      
\end{split}
\end{equation}
Here $\epsilon_{\boldsymbol{k}-}$ is the lower helicity band dispersion, the prime symbol in the second line means the summation is constrained by the condition $\boldsymbol{k_{1}}+\boldsymbol{k_{2}}=\boldsymbol{k_{3}}+\boldsymbol{k_4}$, 
and 
\begin{equation}
\begin{split}
f_{\boldsymbol{k_{1},k_{2}}}^{\boldsymbol{k_{3},k_{4}}}= & \sum_{\sigma\sigma^{\prime}}g_{\sigma\sigma^{\prime}}\langle\boldsymbol{k_{1}}-|\boldsymbol{k_{1}}\sigma\rangle\langle\boldsymbol{k_{2}}-|\boldsymbol{k_{2}}\sigma^{\prime}\rangle \\
& \times\langle \boldsymbol{k_{3}}\sigma^{\prime}|\boldsymbol{k_{3}}-\rangle\langle\boldsymbol{k_{4}}\sigma|\boldsymbol{k_{4}}-\rangle ,
\end{split}
\end{equation}
is the interaction projected to the lower helicity branch.

The low-energy excitation spectrum of the projected Hamiltonian can be obtained by the standard Bogoliubov theory. Assuming the BEC is condensed at the momentum $+\boldsymbol{k}_{m}$, the Bogoliubov-de Gennes (BdG) equation reads
\begin{equation}
\setlength{\arraycolsep}{0.1pt}
H_{\textrm{BdG}}= \dfrac{1}{2}\sum_{\boldsymbol{q}\neq0}\Psi_{\boldsymbol{q}}^{\dagger}K(\boldsymbol{q})
\Psi_{\boldsymbol{q}},
\end{equation}
where $\Psi_{\boldsymbol{q}}^{\dagger}=(a_{\mathbf{k}_{m}+\mathbf{q},-}^{\dagger},a_{\mathbf{k}_{m}-\mathbf{q},-})$ and the Bogoliubov matrix is
\begin{equation}
K(\boldsymbol{q})=
\begin{pmatrix}
  K_{0}(\boldsymbol{q}) +\Sigma_{11}(\boldsymbol{q}) &  \Sigma_{12}(\boldsymbol{q}) \\[3pt]
   \Sigma_{21}(\boldsymbol{q})  &   K_{0}(-\boldsymbol{q}) +\Sigma_{22}(\boldsymbol{q})
\end{pmatrix}. 
\end{equation}
Here $K_{0}(\boldsymbol{q})=\epsilon_{\mathbf{k}_{m}+\mathbf{q},-} -\mu$, and
$\mu=\epsilon_{\boldsymbol{k}_m,-}+\Sigma_{11}(0)-\Sigma_{12}(0)$ is the chemical potential. 
$\Sigma_{ij}$ is the $ij$ component of the self-energy 
\begin{equation}
\begin{aligned} 
\Sigma_{11}(\boldsymbol{q}) = & \frac{n_{0}}{2}\left(f_{\mathbf{k}_{m}, \mathbf{k}_{m}+\mathbf{q}}^{\mathbf{k}_{m}, \mathbf{k}_{m}+\mathbf{q}}+f_{\mathbf{k}_{m}+\mathbf{q}, \mathbf{k}_{m}}^{\mathbf{k}_{m}+\mathbf{q}, \mathbf{k}_{m}}\right. \\ & \left.+f_{\mathbf{k}_{m}+\mathbf{q}, \mathbf{k}_{m}}^{\mathbf{k}_{m}, \mathbf{k}_{m}+\mathbf{q}}+f_{\mathbf{k}_{m}, \mathbf{k}_{m}+\mathbf{q}}^{\mathbf{k}_{m}+\mathbf{q}, \mathbf{k}_{m}}\right), \\ 
\Sigma_{12}(\boldsymbol{q}) = & \frac{n_{0}}{2}\left(f_{\mathbf{k}_{m}+\mathbf{q}, \mathbf{k}_{m}-\mathbf{q}}^{\mathbf{k}_{m}, \mathbf{k}_{m}}+f_{\mathbf{k}_{m}-\mathbf{q}, \mathbf{k}_{m}+\mathbf{q}}^{\mathbf{k}_{m}, \mathbf{k}_{m}}\right),
\end{aligned}
\end{equation}
and other terms of the self-energy can be obtained via $\Sigma_{11}(\boldsymbol{q})=\Sigma_{22}(\boldsymbol{-q})$ and $\Sigma_{12}(\boldsymbol{q})=\Sigma_{21}(\boldsymbol{q})^{\ast}$. $n_0$ is the average condensate density.
The BdG Hamiltonian can be diagonalized by defining the quasiparticle operators
$\alpha_{\boldsymbol{q}}$ and $\alpha_{\boldsymbol{-q}}^{\dagger}$, related with $\Psi_{\boldsymbol{q}}$ by
\begin{equation}
\Psi_{\boldsymbol{q}}\equiv{a_{\boldsymbol{k_{m}+q},-} \choose a_{\boldsymbol{k_{m}-q},-}^{\dagger}}=
\left( 
\begin{array}{cc}
 u_{\boldsymbol{q}} & v_{\boldsymbol{-q}} \\
v_{\boldsymbol{q}}^{\ast} & u_{\boldsymbol{-q}}^{\ast}
\end{array}   
\right)
{\alpha_{\boldsymbol{q}} \choose \alpha_{\boldsymbol{-q}}^{\dagger}},
\end{equation}
where the quasiparticle amplitudes $u_{\boldsymbol{q}}, v_{\boldsymbol{q}}$ and Bogoliubov excitation spectrum $E_{\boldsymbol{q}}$ can be obtained by diagonalizing the matrix $\sigma_z K(\boldsymbol{q})$ \cite{subacsi2022quantum,Zhu_2015,xia2023,chen2023}.
In terms of the quasiparticle operators $\alpha_{\boldsymbol{q}}$, the BdG Hamiltonian now takes the form 
\begin{equation}
H_{\text{BdG}} \simeq \sum_{\boldsymbol{q}}E_{\boldsymbol{q}}\alpha_{\boldsymbol{q}}^{\dagger}\alpha_{\boldsymbol{q}}.
\end{equation}
The excitation spectrum $E_{\boldsymbol{q}}$ obtained in this way agrees well with the lower branch of the excitation spectrum without projection \cite{Zhaihui2013}, as shown in Fig. \ref{fig1}.

For comparison, we also plot the two branches of Bogoliubov excitations by solving the full BdG equation of spin-orbit coupled BEC without projection. As shown in Fig. \ref{fig1}, the excitation spectrum by the projection method agrees well with the lower branch of excitation without projection, validating the reliability of the projection method.

\section{induced anisotropic interactions}
\label{interaction}
It is pointed out that for an impurity immersed in the spin-orbit coupled BEC, the asymmetric Bogoliubov excitation gives rise to a polaron with asymmetric dispersion \cite{Huhui2019}. To analyze the induced interaction between polarons and the bipolarons, it is necessary to first calculate the single polaron dispersion.

\begin{figure}[t]
	\centering
	\includegraphics[width=0.98\linewidth]{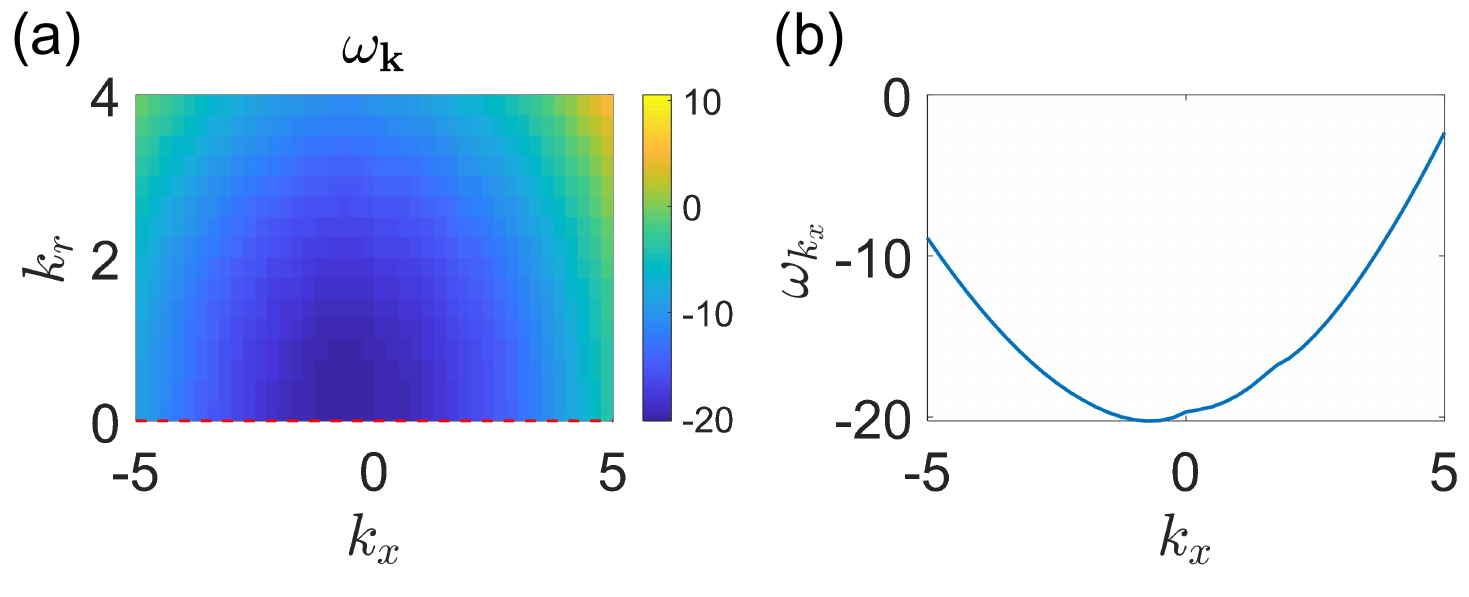}
  \caption{(a) The single polaron dispersion $\omega_{\boldsymbol{k}}$ with the parameter $1/a_{\text{IB}}=-1$. (b) The single polaron dispersion along the  $k_x$ direction (along the red dashed line in (a), i.e., $k_r=0$).}
    \label{fig2}
\end{figure}

The property of a single polaron can be obtained from its Green's function
\begin{equation}
G(i\omega,\boldsymbol{k})=\dfrac{1}{i\omega-\varepsilon_{\boldsymbol{k}}-\Sigma(i\omega,\boldsymbol{k})},
\end{equation}
where $\Sigma(i\omega,\boldsymbol{k})$ is the self-energy due to interaction with BEC and $\varepsilon_{\boldsymbol{k}}=k^2/2m_I$ is the dispersion of free impurity.
The poles of the Green's function give the energy of the quasiparticles $\omega_{\boldsymbol{k}}$, i.e.,
\begin{equation}
\omega_{\boldsymbol{k}}-\varepsilon_{\boldsymbol{k}}-\text{Re}\Sigma(\omega_{\boldsymbol{k}}+i\eta,\boldsymbol{k})=0,
\label{dispersion}
\end{equation}
where $\eta$ is a positive infinitesimal number, and the corresponding quasiparticle residue is
\begin{equation}
  Z_{\boldsymbol{k}}=\dfrac{1}{1-\partial_{\omega}\text{Re}\Sigma(\omega+i\eta,\boldsymbol{k})}\bigg|_{\omega=\omega_{\boldsymbol{k}}}.
\end{equation}
We use the ladder approximation to calculate the impurity self-energy $\Sigma(i\omega,\boldsymbol{k})$, which turns out to be a good approximation in agreement with quantum Monte Carlo results \cite{rath_fieldtheory2013,camacho2018bipolarons,camacho2018landau}. In this approximation, the self-energy is \cite{Huhui2019,dingpolarons2023}
\begin{equation}
\Sigma(i\omega,\boldsymbol{k})=n_{0}\mathcal{T}(i\omega,\boldsymbol{k}),
\end{equation}
with the impurity-boson scattering matrix $\mathcal{T}(i\omega,\boldsymbol{k})$ given by 
\begin{equation}
\mathcal{T}(i\omega,\boldsymbol{k})=\dfrac{1}{g_{\text{IB}}^{-1}+\Pi(i\omega,\boldsymbol{k})}.
\end{equation}
Here $g_{\text{IB}}=2\pi a_{\text{IB}}/{m_{\mu}}$ is the interaction between impurities and bosons, and $a_{\text{IB}}$ is the scattering length between the impurities and bosons, which is assumed to be independent of the pseudospin of bosons. $m_{\mu}=m_{B}m_{I}/(m_{B}+m_{I})$ is the reduced mass, and $\Pi(i\omega,\boldsymbol{k})$
is the renormalized pair propagator. After performing the summation of Matsubara frequencies, the expression of pair propagator reads \cite{rath_fieldtheory2013},
\begin{equation}
\Pi(i\omega,\boldsymbol{k})=-\int\dfrac{d^3\boldsymbol{q}}{(2\pi)^{3}}\left(\dfrac{u_{\boldsymbol{q}}^{2}}{i\omega-E_{\boldsymbol{q}}-\varepsilon_{\boldsymbol{k-q}}}+\dfrac{2m_{\mu}}{\boldsymbol{q}^{2}}\right),
\end{equation}
where $E_{\boldsymbol{q}}$ is the anisotropic Bogoliubov excitation spectrum of spin-orbit coupled BEC. We have
chosen $m_{B}=m_{I}=m$ throughout this paper for simplicity. The pair propagator can only be calculated numerically for the anisotropic case. In practice, we have performed the numerical integration in cylindrical coordinate, since the system has rotational symmetry along the $x$ direction.

With the pair propagator, the single polaron dispersion can be obtained by numerically solving Eq. \ref{dispersion} and the result is shown in Fig. \ref{fig2}. Note that there are two branches of polaron dispersion, and we only focus on the attractive branch, whose energy is negative and represents a well-defined quasiparticle. As shown in Fig. \ref{fig2}, the single polaron dispersion is asymmetric, i.e., $\omega_{\boldsymbol{k}}\neq\omega_{-\boldsymbol{k}}$. In addition, the minimum of the dispersion is located at nonzero momentum, due to interaction with the asymmetric Bogoliubov excitations, which is in agreement with previous findings \cite{Huhui2019}. Note that the momentum at the dispersion minimum is close to zero rather than near the roton minimum, indicating that the polaron in the parameter regime considered here remains phonon-induced rather than roton-induced \cite{Huhui2019}.

\begin{figure}[tb]
\centering
\includegraphics[width=0.98\linewidth]{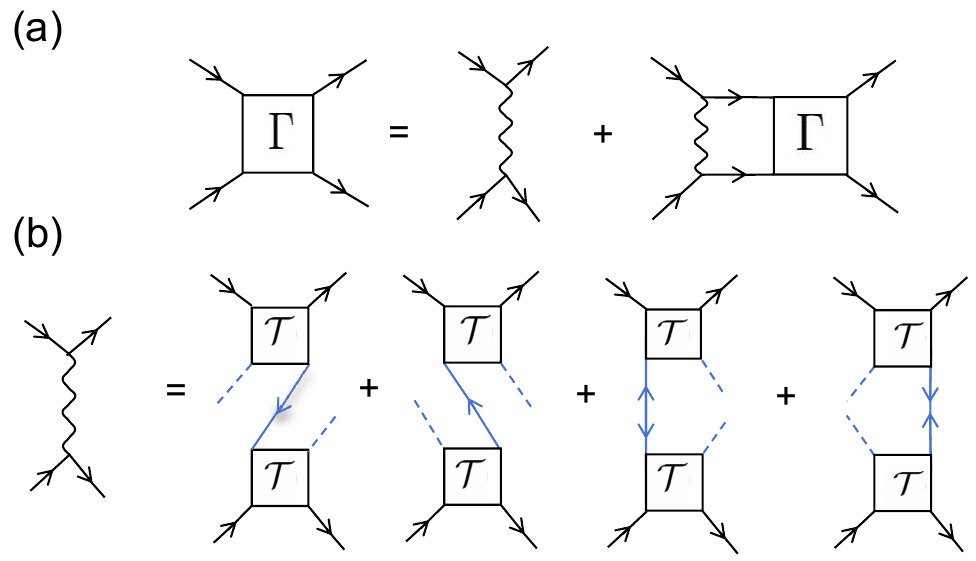}
 \caption{(a) Diagrammatic representation of the Bethe-Salpeter equation. The black lines are the impurity Green’s function, and the wavy line is the induced interaction. (b) The diagram for the induced interaction. The solid blue lines are the normal and anomalous Green’s functions of spin-orbit coupled BEC, and the dashed
blue lines represent condensate bosons.}
 \label{fig3}
\end{figure}

In order to derive the effective interaction between polarons, we start from the scattering matrix $\Gamma(k_{1},k_{2};q)$, which describes the scattering of two impurities from energy momenta ($k_{1}$, $k_{2}$)  to ($k_{1}+q$, $k_{2}-q$), with $k=(z,\boldsymbol{k})$ being the four-momentum vector. The scattering matrix obeys the Bethe-Salpeter equation, which in the ladder approximation reads (see Fig. \ref{fig3}(a)) \cite{fetter2012quantum,altland2010condensed}
\begin{equation} \label{e:BSE}
\begin{split}
\Gamma(k_{1},k_{2};q)  = & V(k_{1},k_{2};q)+\sum_{q^{\prime}}
V(k_{1},k_{2};q^{\prime})\\ & \times G(k_{1}-q^{\prime}) G(k_{2}+q^{\prime})\\
 & \times\Gamma(k_{1}-q^{\prime},k_{2}+q^{\prime};q-q^{\prime}),
\end{split}
\end{equation}
where $V(k_{1},k_{2};q)$ is the induced interaction between the impurities.
A direct solution of the Bethe-Salpeter equation is very challenging, and in order to simplify the problem, we therefore makes some approximations outlined in Refs. \cite{camacho2018bipolarons,dingpolarons2023}. First, we make a pole expansion for the impurity's Green's function, i.e., $G(k)\simeq Z_{\boldsymbol{k}}/(i\omega-\omega_{\boldsymbol{k}})$. To transform from the picture of bare impurity to the quasiparticle polaron, we multiply the equation by $Z_{\boldsymbol{k_{1}}}Z_{\boldsymbol{k_{2}}}$. This gives an effective polaron-polaron interaction 
$V_{\text{eff}}(k_{1},k_{2};q)= Z_{\boldsymbol{k_{1}}}Z_{\boldsymbol{k_{2}}} V(k_{1},k_{2};q)$
and the scattering matrix between two polarons is
$\Gamma_{\rm{polaron}}(k_{1},k_{2};q)= Z_{\boldsymbol{k_{1}}}Z_{\boldsymbol{k_{2}}} \Gamma(k_{1},k_{2};q)$.
Second, we neglect the frequency dependence of the induced interaction and take the static limit, which is reasonable when the binding energy of the bipolaron state is small compared to the typical energy of the Bogoliubov excitations. These approximations are good enough as long as the quasiparticles are well defined and the binding energy of bipolaron is small. In this case, the frequency summation in Eq. \ref{e:BSE} can be performed analytically, and then the Bethe-Salpeter equation is just reduced to the Lippmann-Schwinger equation, which is equivalent to an effective Schrödinger equation for two polarons interacting via an induced interaction (see Eq. \ref{eq:sc}).

With above approximations, the effective interaction between two polarons scattered from momenta ($\boldsymbol{k}, -\boldsymbol{k}$) to ($\boldsymbol{k^{\prime}}, -\boldsymbol{k^\prime}$), as illustrated by the diagrams in Fig. \ref{fig3}(b), is simplified to (in the
center-of-mass frame) \cite{camacho2018bipolarons,dingpolarons2023}
\begin{equation}\label{eq:INDint}
\begin{split}
V_{\text{eff}}(\boldsymbol{k,k^{\prime}})&= Z_{\boldsymbol{k}}Z_{-\boldsymbol{k}}n_{0}
[\mathcal{T}(\omega,\boldsymbol{k})G_{11}(0,\boldsymbol{k-k^{\prime}})\mathcal{T}(\omega,\boldsymbol{-k^{\prime}}) \\
  &\quad+\mathcal{T}(\omega,\boldsymbol{k^{\prime}})G_{11}(0,\boldsymbol{k^{\prime}-k})\mathcal{T}(\omega,\boldsymbol{-k}) \\                            &\quad+\mathcal{T}(\omega,\boldsymbol{k^{\prime}})G_{12}(0,\boldsymbol{k^{\prime}-k})\mathcal{T}(\omega,\boldsymbol{-k^{\prime}})\\   &\quad+\mathcal{T}(\omega,\boldsymbol{k})G_{21}(0,\boldsymbol{k^{\prime}-k})\mathcal{T}(\omega,-\boldsymbol{k})],
\end{split}
\end{equation}
where $G_{ij}(0,\boldsymbol{k})$ is the Green's function for spin-orbit coupled BEC after band projection,
\begin{equation}
G_{11}(i\omega,\boldsymbol{k})=\dfrac{|u_{\boldsymbol{k}}|^{2}}{i\omega-E_{\boldsymbol{k}}}   - \dfrac{|v_{\boldsymbol{-k}}|^{2}}{i\omega+E_{\boldsymbol{-k}}},
\end{equation}
\begin{equation}
G_{12}(i\omega,\boldsymbol{k})=\dfrac{ u_{\boldsymbol{k}}v_{\boldsymbol{k}}}{i\omega-E_{\boldsymbol{k}}}   - \dfrac{u_{\boldsymbol{-k}}v_{\boldsymbol{-k}}}{i\omega+E_{\boldsymbol{-k}}},
\end{equation}
and other terms are related via 
\begin{equation}
\begin{split}
   G_{11}(i\omega,\boldsymbol{k})= &  G_{22}(-i\omega,\boldsymbol{-k}),\\
   G_{12}(i\omega,\boldsymbol{k})= &  G_{21}(-i\omega,\boldsymbol{-k}).
   \end{split}
\end{equation}
Here $\omega$ is taken to be the energy of the interacting quasiparticles.

\begin{figure}[!tb]
  \centering
    \includegraphics[width=0.98\linewidth]{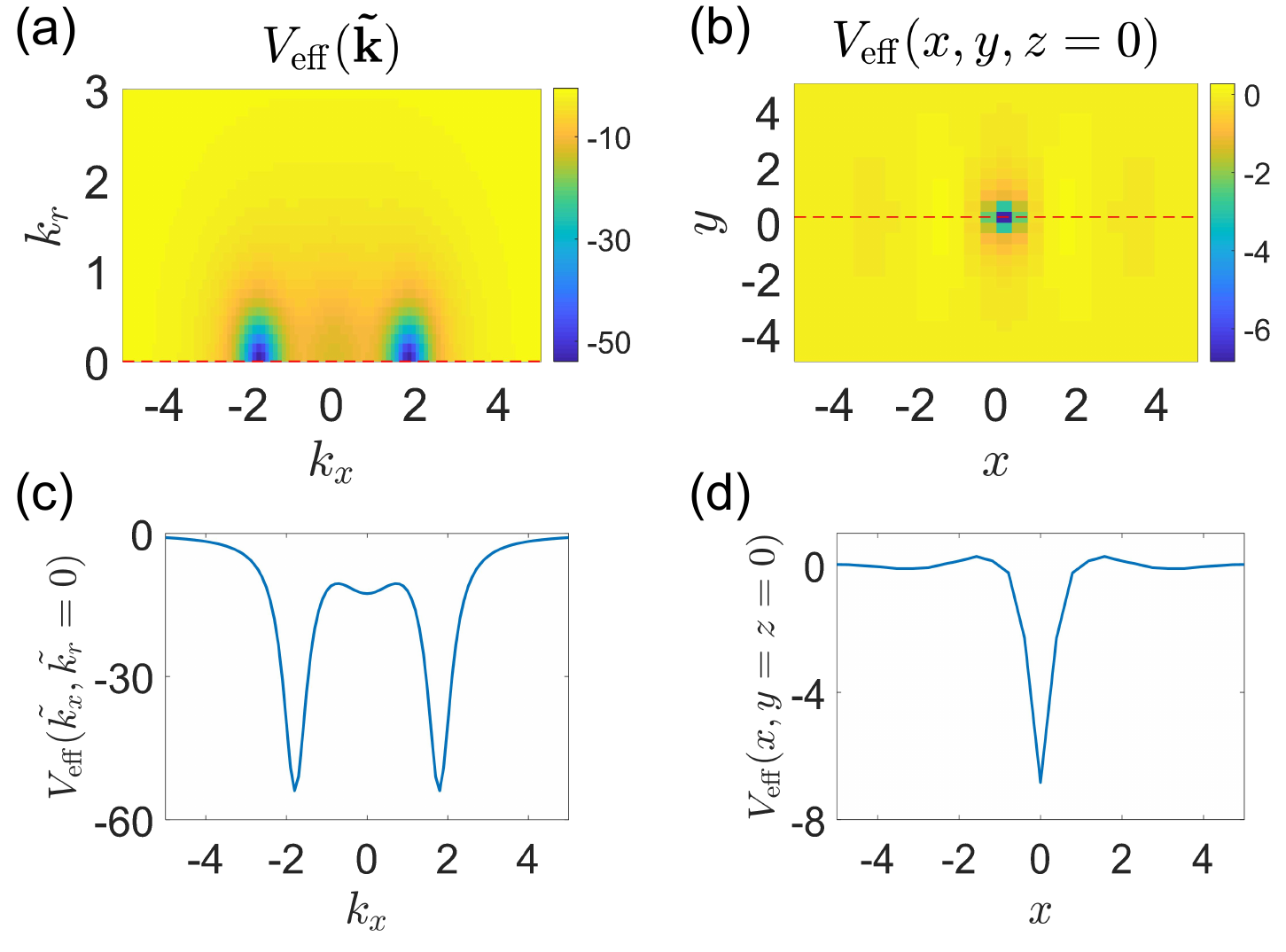}
    \caption{(a) The induced interaction $V_{\text{eff}}(\boldsymbol{k-k^{\prime}})\equiv V_{\text{eff}}(\tilde{\boldsymbol{k}})$ as a function of $k_r$ and $k_x$ with $a_{\text{IB}}=-0.1$. (b) $V_{\text{eff}}(x,y,z=0)$ in the real space. 
    (c) $V_{\text{eff}}(\tilde{k}_x,\tilde{k}_r=0)$ and (d) $V_{\text{eff}}(x,y=z=0)$ are line cuts of the 2D plots in (a) and (b), respectively, along the red dashed lines.}
    \label{fig4}
\end{figure}

For weak impurity-boson interaction, the scattering matrix can be reduced to $\mathcal{T}(\omega,\boldsymbol{k})\simeq g_{\text{IB}}=2\pi a_{\text{IB}}/m_{\mu}$, and the effective interaction is now 
\begin{equation}\label{eq:weak}
\begin{split}
 V_{\text{eff}}(\boldsymbol{k-k^{\prime}})  \equiv  &  V_{\text{eff}}(\tilde{\boldsymbol{k}}) \\
  =  &  n_{0}g_{\text{IB}}^{2}[G_{11}(0,\tilde{\boldsymbol{k}})+G_{12}(0,\tilde{\boldsymbol{k}})   \\
  & + G_{11}(0,-\tilde{\boldsymbol{k}})+G_{12}(0,-\tilde{\boldsymbol{k}})].  
\end{split}
\end{equation}
 This result agrees with the second-order perturbation theory, which is applicable only when the impurity-boson interaction is weak \cite{camacho2018bipolarons,camacho2018landau}. In Fig. \ref{fig4}, we plot the profile of $V_{\text{eff}}(\tilde{\boldsymbol{k}})$ in Eq. \ref{eq:weak} in the momentum and real space, respectively. In momentum space, it has two symmetric double minima in the $k_x$ direction, and after careful inspection, the minima turn out to be related with the roton minimum in the Bogoliubov excitation spectrum of spin-orbit coupled BEC. The momenta at the minima of $V_{\text{eff}}(\tilde{\boldsymbol{k}})$ coincide with the momentum at the minimum of roton excitation. This indicates that the interaction is roton-induced, in contrast to the conventional phonon-induced interaction with minimum at zero momentum \cite{camacho2018bipolarons,camacho2018landau,viverit2000zero,yu2012induced}. In addition, the effective interaction is also notably anisotropic, a characteristic generally expected for any BEC with anisotropic Bogoliubov excitations. By the Fourier transform, we also show the effective interaction in real space, which has an oscillation feature along the $x$ direction due to the double minima feature in momentum space, in clear contrast to the Yukawa-type interaction for the case of isotropic phonons. 

\begin{figure}[tb]
	\centering
	\includegraphics[width=0.9\linewidth]{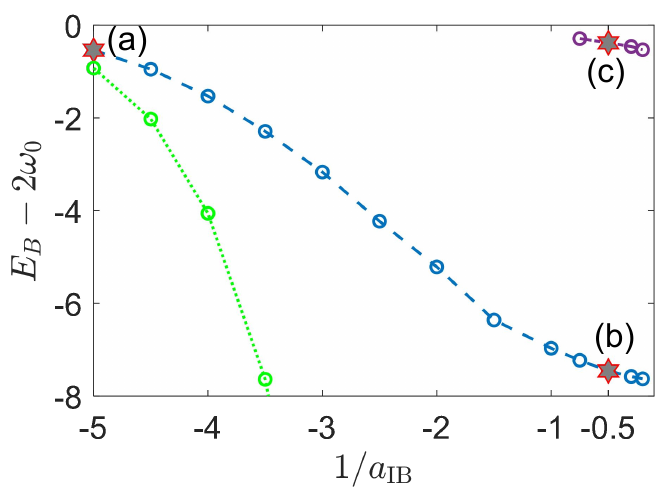}
  \caption{Binding energy $E_{B}$ with respect to $2 \omega_{0}$ ($2 \omega_{\boldsymbol{k}=0}$) as a function of $1/a_{\text{IB}}$. The blue line and purple line are the binding energies obtained by solving Eq. \ref{eq:sc} with the induced interaction given by Eq. \ref{eq:INDint}, and the green line gives the binding energy with the induced interaction given by Eq. \ref{eq:weak}, i.e., the weak impurity-boson interaction limit. }
  \label{fig5}
\end{figure}
 
\section{bipolarons}
\label{bipolaron}
With the single polaron dispersion $\omega_{\boldsymbol{k}}$ in Eq. \ref{dispersion} and effective interaction between polarons $V_{\text{eff}}(\boldsymbol{k},\boldsymbol{k^{\prime}})$ in Eq. \ref{eq:INDint}, we are now ready to solve the bipolaron energies and wave functions. Taking the center-of-mass momentum of the two interacting polarons to be zero, the effective Schrödinger equation in the momentum space is expressed as follows,
\begin{equation}\label{eq:sc}
(\omega_{\boldsymbol{k}}+\omega_{\boldsymbol{-k}})\psi(\boldsymbol{k})+\sum_{\boldsymbol{k^{\prime}}}V_{\text{eff}}(\boldsymbol{k},\boldsymbol{k^{\prime}})\psi(\boldsymbol{k^{\prime}})=E_{B}\psi(\boldsymbol{k}),
\end{equation}
where $\psi(\boldsymbol{k})$ is the wave function for the relative momentum of the bipolaron and $E_{B}$ is the binding energy. $V_{\text{eff}}(\boldsymbol{k},\boldsymbol{k^{\prime}})$ is the interaction matrix for two polarons scattered from momenta ($\boldsymbol{k}, -\boldsymbol{k}$) to ($\boldsymbol{k^{\prime}}, -\boldsymbol{k^\prime}$). Note that even for asymmetric Bogoliubov excitation $E_{\boldsymbol{k}}\neq E_{-\boldsymbol{k}}$ and single polaron dispersion $\omega_{\boldsymbol{k}}\neq\omega_{\boldsymbol{-k}}$, the effective interaction always satisfies $V_{\text{eff}}(\boldsymbol{k},\boldsymbol{k^{\prime}})=V_{\text{eff}}(\boldsymbol{k^{\prime}},\boldsymbol{k})$. This symmetry has guaranteed that the Hamiltonian for Eq. \ref{eq:sc} is always hermitian, and thus the binding energy $E_B$ is always real.

\begin{figure}[!t]
\centering
\includegraphics[width=0.98\linewidth]{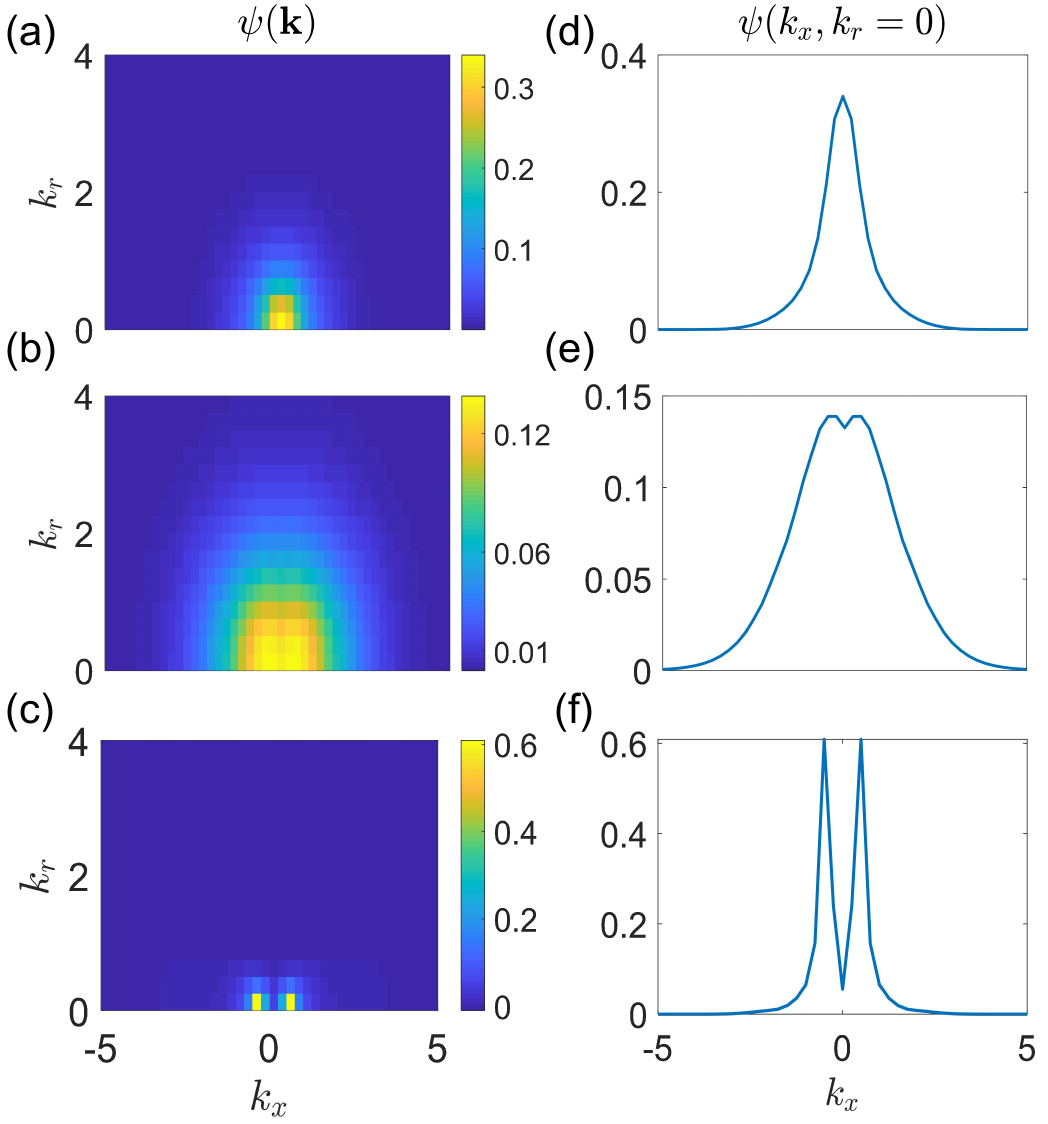}
 \caption{Bipolaron wave functions in the momentum space for different impurity-boson interactions. (a) for $1/a_{\text{IB}}=-5$, and (b), (c) for $1/a_{\text{IB}}=-0.5$. These three wave functions correspond to the parameters indicated by red stars with same label in Fig. \ref{fig5}. (d), (e) and (f) are line cuts of the 2D plots in (a), (b) and (c), respectively.}
 \label{fig6}
\end{figure}

We solve Eq. \ref{eq:sc} numerically in cylindrical coordinate, and plot the binding energy $E_B$ as a function of impurity-boson scattering length $a_{\text{IB}}$ in Fig. \ref{fig5}. As can be seen from the figure, the solutions with $V_{\text{eff}}(\boldsymbol{k},\boldsymbol{k^{\prime}})$ in Eq. \ref{eq:INDint} and Eq. \ref{eq:weak} match only when the impurity-boson interaction is weak enough. This illustrates the limitations of second-order perturbation theory and the necessity of using the effective interaction $V_{\text{eff}}(\boldsymbol{k},\boldsymbol{k^{\prime}})$ from Eq. \ref{eq:INDint} in the calculation of bipolaron problem, especially for the case of strong impurity-boson interaction \cite{camacho2018bipolarons}. In addition, when the impurity-boson interaction exceeds a critical value, there exist two bipolarons.

In Fig. \ref{fig6}, we plot the corresponding bipolaron wave functions, for both the weak and strong impurity-boson interactions. It is found that when the impurity-boson interaction is weak, the bipolaron wave function has similar feature with conventional isotropic case. The maximum of the wave function in momentum space resides at zero momentum and decays with increase of $k$. In contrast, when the impurity-boson interaction is strong, the bipolaron wave function exhibits much different feature, i.e., its maxima are now located at nonzero momenta instead, which is more clear from the line cut along the $x$ direction. The wave function at higher energy also has this ``double-peak'' structure, and deviation from the isotropic case is much more pronounced. This feature can be understood from the fact that
when the impurity-boson interaction is strong, the single polaron dispersion $\omega_{\boldsymbol{k}}$ becomes more asymmetric in the $k_x$ direction and the minimum of the polaron energy is located at a nonzero momentum \cite{Huhui2019}. The rough agreement between the momentum at the dispersion minimum in Fig. \ref{fig2}(b) and the momenta at the double peak in Fig. \ref{fig6} confirms the validity of this interpretation. Note that all these wave functions are inversion symmetric with respect to $\boldsymbol{k}\leftrightarrow-\boldsymbol{k}$ in the parameter regime considered here, which means these wave functions are only relevant for bosonic impurities.

\section{summary}
\label{conclusion}
In summary, we have studied the properties of the induced interaction between polarons, mediated by the exchange of anisotropic Bogoliubov excitations of a BEC, as well as the characteristics of bipolarons arising from this interaction. Using spin-orbit coupled BEC as an example, where the Bogoliubov excitations are both anisotropic and asymmetric, we find that the induced interaction is also anisotropic. Furthermore, the roton excitation spectrum introduces a distinctive double-minima feature in the induced interaction in momentum space, in stark contrast to the conventional interaction arising from isotropic excitations. The resulting bipolaron wave functions are also anisotropic, and for strong impurity-boson interaction, display a double-peak feature. Given the experimental realization of 1D SOC in BEC, the unique properties of roton-induced interactions and new features in bipolarons predicted in this work can be tested using existing experimental techniques in the near future.

\begin{acknowledgements}
Z.Y. and Q.Z. are supported by the National Key Research and Development Program of China (Grant No. 2022YFA1405304), the National Natural Science Foundation of China (Grant No. 12004118) and the Open Fund of Key Laboratory of Atomic and Subatomic Structure and Quantum Control (Ministry of Education). S.D. is supported by the Fundamental Research Funds for the Central Universities and the National Natural Science Foundation of China (Grant No. 12404320).
\end{acknowledgements}

\end{document}